\documentclass[10pt]{amsart}

\usepackage{amsmath}
\usepackage{graphicx}
\usepackage{cite}

\usepackage{amsbsy}
\usepackage{subfig}
\usepackage{soul}

\title{Analogue Hawking radiation in an exactly solvable model of BEC}
\author{Alberto Parola$^{1}$ \and Manuele Tettamanti$^{1,2}$ \and Sergio L. Cacciatori $^{1,3}$}

\address{\noindent $^1$Dipartimento di Scienza e Alta Tecnologia, Universit\`a dell' Insubria, via Valleggio 11, 22100 Como, Italy \endgraf
$^2$INO-CNR BEC Center, I-38123 Povo, Italy \endgraf
$^3$INFN Sezione di Milano, via Celoria 16, 20133 Milano, Italy}

\begin{document}

\begin{abstract}{
Hawking radiation, the spontaneous emission of thermal photons from an event horizon, is one of the most intriguing and elusive predictions of field theory in curved spacetimes. A formally analogue phenomenon occurs at the supersonic transition of a fluid: in this respect, ultracold gases stand out among the most promising systems but the theoretical modelling of this effect has always been carried out in semiclassical approximation, borrowing part of the analysis from the gravitational analogy. Here we discuss the exact solution of a one-dimensional Bose gas flowing against an obstacle, showing that spontaneous phonon emission (the analogue of Hawking radiation) is predicted without reference to the gravitational analogy. Long after the creation of the obstacle, the fluid settles into a stationary state displaying the emission of sound waves (phonons) in the upstream direction. A careful analysis shows that a precise correspondence between this phenomenon and the spontaneous emission of radiation from an event horizon requires additional conditions to be met in future experiments aimed at identifying the occurrence of the Hawking-like mechanism in Bose-Einstein condensates.}
\end{abstract}

\maketitle

The spontaneous emission of thermal radiation from an event horizon (the so-called ``Hawking radiation'') is one of the most intriguing consequences of quantum field theory in curved spacetimes. Soon after the original prediction on black holes~\cite{hawking1,hawking2}, the occurrence of a similar effect in other frameworks (fluid dynamics, optics, ultra-cold gases) was suggested~\cite{unruh}, stimulating an intense quest for experimental and theoretical evidence of ``analogue gravity'' phenomena~\cite{comoschool}. In the context of fluids, the formation of a ``sonic horizon'', with features quite similar to a black-hole horizon, is a known hydrodynamical effect occurring at the subsonic-supersonic transition of either a classical or a quantum fluid~\cite{unruh} and, according to the analogue gravity paradigm, sound waves should originate from the vacuum fluctuations of the phonon field at the sonic transition~\cite{liberati}. Following the gravitational analogy, the phonon spectrum is predicted to be thermal with an effective temperature $T_H$ related to the gradient of the velocity field at the horizon, i.e. at the point $x_h$ where the fluid velocity $v(x_h)$ equals the local sound velocity $c_s(x_h)$:

\begin{equation}
k_BT_H = \frac{\hbar}{2\pi}\frac{d(v-c_s)}{dx}\Bigg\vert_{x=x_h} .
\label{unruh}
\end{equation} 

In the past years, Bose-Einstein condensates (BECs) were considered as possible candidates \cite{garay1,garay2,fischer1,fischer2} and an experimental realization of a sonic black hole in a flowing BEC was first achieved few years ago \cite{stein2010}. More recently, two experiments \cite{stein14,stein16} made a considerable progress towards the direct observation of Hawking radiation in a quasi-one-dimensional set-up but the complete characterization of the observed phenomena is still under debate \cite{tettamanti,ted1,ted2,stein-nova,ulf,stein-risp}.
In Bose-Einstein condensates this mechanism has always been studied within semiclassical approximation via the Gross-Pitaevskii equation supplemented by a Bogoliubov analysis of the excitation spectrum (see, for instance, ~\cite{comoschool,parentani,parentani1,larre,recati}), 
while a seminal investigation in Fermi gases related the phonon emission at the sonic horizon to the quantum reflection against a potential barrier~\cite{giovanazzi}. Nevertheless, up to date a fully microscopic model displaying the analogue gravity mechanism at work in a physically well-defined setting has not been devised yet, even though it has been deemed desirable soon after the discovery of the analogy~\cite{unruh1}. Such an investigation would allow to understand the physical origin of the quasi-particle emission in a condensed matter system and to establish the necessary conditions for the detection and the characterization of this effect. This is therefore a crucial step towards the unambiguous identification of the analogue Hawking radiation in a laboratory. 

In this Letter we present the exact solution of a model of a one-dimensional Bose gas flowing in an external potential. In particular, the collapse of a star into a final black hole is simulated by a quantum quench perturbing a free stationary flow. We then analyse the sound spectrum at late times, when the stationary regime is met again. Sound waves propagating in the upstream direction are always generated, but they acquire the thermal character typical of the Hawking radiation only if further conditions are met. 
Besides the mandatory presence of a sonic horizon \cite{unruh}, we find that a faithful correspondence with the Hawking mechanism requires a careful choice of the external potential, which must be an extremely smooth barrier with height 
smaller than the largest kinetic energy of the fluid particles. Surprisingly, it implies that other popular choices, like steep barriers or ``waterfall" potentials 
(often used in the experiments, e.g.,~\cite{stein14,stein16}), are not appropriate in the analogue gravity framework. These conditions are strictly related to the condensed-matter nature of the system and they emerge as new requirements for the occurrence of the analogue Hawking effect in BECs. To the best of our knowledge, this is the first work where an exact analysis of the analogue Hawking radiation in Bose-Einstein condensates is presented.

\section{The model}

We study a one-dimensional fluid of point-like hard-core bosons (HCB) at zero physical temperature, also known as Tonks-Girardeau gas~\cite{girardeau}. This particular system was realized in a laboratory few years ago by use of $^{87}$Rb atoms in elongated quasi-1D traps \cite{kinoshita, bloch}. The dynamics of the HCB gas is known to be well represented by the Gross-Pitaevskii equation (GPE)\footnote{More precisely, a slight modification to the GPE is required to provide a fully satisfactory description of the Tonks-Girardeau limit \cite{menotti,salasnich}. Nevertheless, the analogue metric can be recovered in this case also~\cite{cpt}.}~\cite{stringari}. On this basis, it is possible to build the gravitational analogy by showing that phonons couple to an effective acoustic metric which, in principle, can display the presence of an analogue black-hole horizon~\cite{garay1}. A semiclassical study of the Tonks-Girardeau gas in the analogue gravity framework was carried out along these lines in Ref.~\cite{giovanazzi}. We proceed as follows: first we set the gas into motion with a constant speed (this represents the vacuum of the gravitational case) and we make the flow hit an obstacle so to obtain supersonic speeds; then, once the system has entered a stationary state, we study the phonons travelling in the upstream direction and reaching infinity: according to the gravitational analogy they correspond to the Hawking radiation of the quantum field, emitted at the subsonic-supersonic transition (the analogue of the black-hole horizon). By exploiting the Tonks-Girardeau limit, we are able to perform the calculation without any approximation; through an exact mapping, in fact, the Hamiltonian, the density operator and the current operator of the HCB gas coincide with those of a non-interacting Fermi system. 
%The energy spectrum and all the $n$-body correlation functions involving the Hamiltonian and the density operators
%are easily obtained because the system maps into a free Fermi gas. 
%Besides the energy spectrum, also all the $n$-body correlation functions involving the Hamiltonian and the 
%density operators are the same as those of the free Fermi gas. 
%These known features of the HCB gas remain unchanged if an external potential is switched on. 
Therefore, in the absence of external potentials, the ground state of a HCB gas is described by a Fermi distribution in the interval $(-k_F,k_F)$, where the Fermi momentum is easily related to the (uniform) particle density by $k_F=\pi\,\rho$. Following the procedure just described, we first perform a Galileo transformation and we shift the wave vectors of all particles by $-k_0$, setting the fluid into motion towards the left with uniform velocity $v_0=\frac{\hbar k_0}{m}$. The fermionic many-body state is a Slater determinant of plane waves $e^{ipx}$ 
with wave vectors in the interval $-k_+ < p < k_-$, where we defined $k_\pm=k_F \pm k_0$.
%$-k_F-k_0 < p < k_F-k_0$. 
In the following we will limit our analysis to the case $k_0 \le k_F$, which corresponds to a subsonic flow. Now we perform a quantum quench: an external potential $V(x)$ is suddenly turned on, bringing the system out of the initial stationary state\footnote{The mapping between the HCB fluid and a Fermi gas remains unchanged if an external potential is switched on \cite{girardeau}.}. As a consequence, each single-particle plane wave evolves in time according to:
%The expectation value of any observable built via {\it local} operators, like the density $\rho(x,t)$ or 
%the current $j(x,t)$, can be expressed in terms of the time evolution of the original plane wave states:
\begin{eqnarray}
\label{evol}
\psi_p(x,t) &=& \int_{-\infty}^{\infty} dk \, c_p(k) \,\phi_k(x)\,e^{-\frac{i}{\hbar}\epsilon_k t} \, ,\\
c_p(k) &=& \int_{-\infty}^{\infty} \frac{dx}{\sqrt{2\pi}} \, \phi^*_k(x)\,e^{ipx}\,e^{-\eta |x|} \, ,
\end{eqnarray}
where $\phi_k(x)$ are the exact single-particle eigenfunctions in the presence of the external potential, $\epsilon_k$ are the corresponding eigenvalues and $\eta\to 0^+$ is the usual convergence factor. Now, let us consider the long-time behaviour of these wave functions {\it fixing the position $x$}. A careful analysis shows that, for each $p$, $\psi_p(x,t)$ is given by the sum of an exact eigenfunction of the system in the presence of the external potential (multiplied by a time dependent phase factor), plus a contribution which vanishes as $t\to+\infty$. These two terms represent, respectively, the asymptotic stationary state and a travelling wave, which originates in the quench and moves to infinity. The asymptotic stationary wave function describing the long-time behaviour of each single-particle state at fixed position $x$ is just the exact scattering eigenstate corresponding to an incident plane wave with wave vector $p$. The evaluation of the asymptotic properties can be made explicit in specific models, where the exact single-particle eigenstates are known, as in the case of a barrier or a ``waterfall" external potential. 

\section{Barrier potential}

In this Letter we will discuss in some detail the case of a barrier defined by: 
\begin{equation}
V(x)=\frac{V_0}{\cosh^2(\alpha x)} \, ,
\label{cosh}
\end{equation}
where $V_0=\frac{\hbar^2Q^2}{2m}$ and $\alpha$ are arbitrary positive parameters defining the height and the 
width of the barrier. 
In this way, $\hbar Q$ represents the momentum of a (classical) particle having just enough energy to overcome the barrier. Potential barriers of similar shape can be experimentally obtained by use of a laser beam which generates an easily tunable repulsive potential~\cite{engels}. The family of external potentials~(\ref{cosh}) is especially convenient because the exact single-particle eigenstates can be expressed in terms of hypergeometric functions~\cite{landau} and the scattering state with wave vector $p>0$ is explicitly given by:
\begin{equation}
\phi_p(x) = \frac{T_p}{\sqrt{2\pi}} \, 
%\frac{\Gamma(a)\Gamma(b)}{\Gamma(c)\Gamma(c-1)}\,
\left [z(1-z)\right]^{-i\frac{p}{2\alpha}}\,
F\left (a, b;c; z\right ) \, ,
\label{eigen}
\end{equation}
where 
$a=\frac{1}{2}-i\frac{p+Q}{\alpha}$, $b=\frac{1}{2}-i\frac{p-Q}{\alpha}$, $c=1-i\frac{p}{\alpha}$ and $z=\frac{1}{2}[1-\tanh(\alpha x)]$; $T_p$ is the transmission amplitude:
\begin{equation}
T_p = \frac{\Gamma(a)\Gamma(b)}{\Gamma(c)\Gamma(c-1)}
\label{tras}
\end{equation}
and $\epsilon_p=\frac{\hbar^2 p^2}{2m}$ is the energy eigenvalue. The case $p<0$ is obtained by replacing $p\to -p$ and $x\to -x$. 
According to the previous analysis, the long-time behaviour of the system is then described by a Slater determinant of eigenstates~(\ref{eigen}) with $-k_+<p<k_-$.
%$-k_F-k_0 < p < k_F-k_0$. 
The model is defined by four parameters with the dimension of an inverse length: $k_F, k_0, Q, \alpha$. 
The expectation value of the current (the mass flux), denoted by $j$, is space-independent and equal to 
\begin{equation}
j = - \hbar \int_{k_-}^{k_+} \frac{dp}{2\pi} \,p\,\vert T_p\vert ^2 \, .
\label{gei}
\end{equation} 
Evaluating the expectation values of local operators in the far upstream region (i.e. $x\to+\infty$) just requires the asymptotic form of the eigenfunctions; the final result involves the reflection coefficient $|R_p|^2=1-|T_p|^2$, leading to expressions identical to a ``quasi-thermal" average defined by the effective momentum distribution:
\begin{equation}
f(p) = \frac{1}{2\pi}\,
\begin{cases} 
1   \,\,\quad -k_+<p< k_-  \\
|R_p|^2  \quad k_-< p<k_+ 
%1   \,\,\quad -k_F-k_0<p< k_F-k_0  \\
%|R_p|^2  \quad k_F-k_0< p<k_F+k_0 
\end{cases} .
\label{effective}
%\nonumber
\end{equation}
This function displays quite a different behaviour at the two Fermi points of the momentum distribution before the quench: while the sharp jump, typical of the zero temperature limit, remains unaltered at $p=-k_+$, $f(p)$ acquires a smooth tail beyond the right Fermi momentum $p=k_-$ up to the cut-off $k_+$. Thus, all the density correlations measured by a hypothetical observer in the upstream region, far away from the barrier, long after the quench, would be indistinguishable from those of a free Fermi fluid with such a momentum distribution. The emerging analogue Hawking radiation is contained 
in this particular form of $f(p)$.

\section{Hawking radiation}

In the gravitational framework the quantized scalar field is defined on a pre-assigned background metric while in analogue models the phonon field represents the excitations of the original quantum system (i.e. the HCB flow) providing the analogue metric. Therefore, a faithful correspondence between the gravitational and the analogue model requires that the quasi-particles of the HCB system behave as an independent, free, quantum scalar field. This crucial condition is verified only in the low ``temperature" limit, i.e., when the difference of the effective momentum distribution from a pure step function  is small~\cite{landau-stat}. 
%In fact, only in this case the excited modes of a Bose gas can be identified with phonons and the non linearities of the dispersion relation are irrelevant. 
This requirement is met provided that the barrier is extremely smooth, i.e., $\frac{\alpha}{Q}\ll 1$: steep external potentials will unavoidably excite high energy modes of the HCB gas, introducing a finite lifetime of the elementary excitations and spoiling the correspondence between the phonon gas and a free quantum field. 

\begin{figure} [h!]
\begin{center}
\includegraphics[scale=0.5]{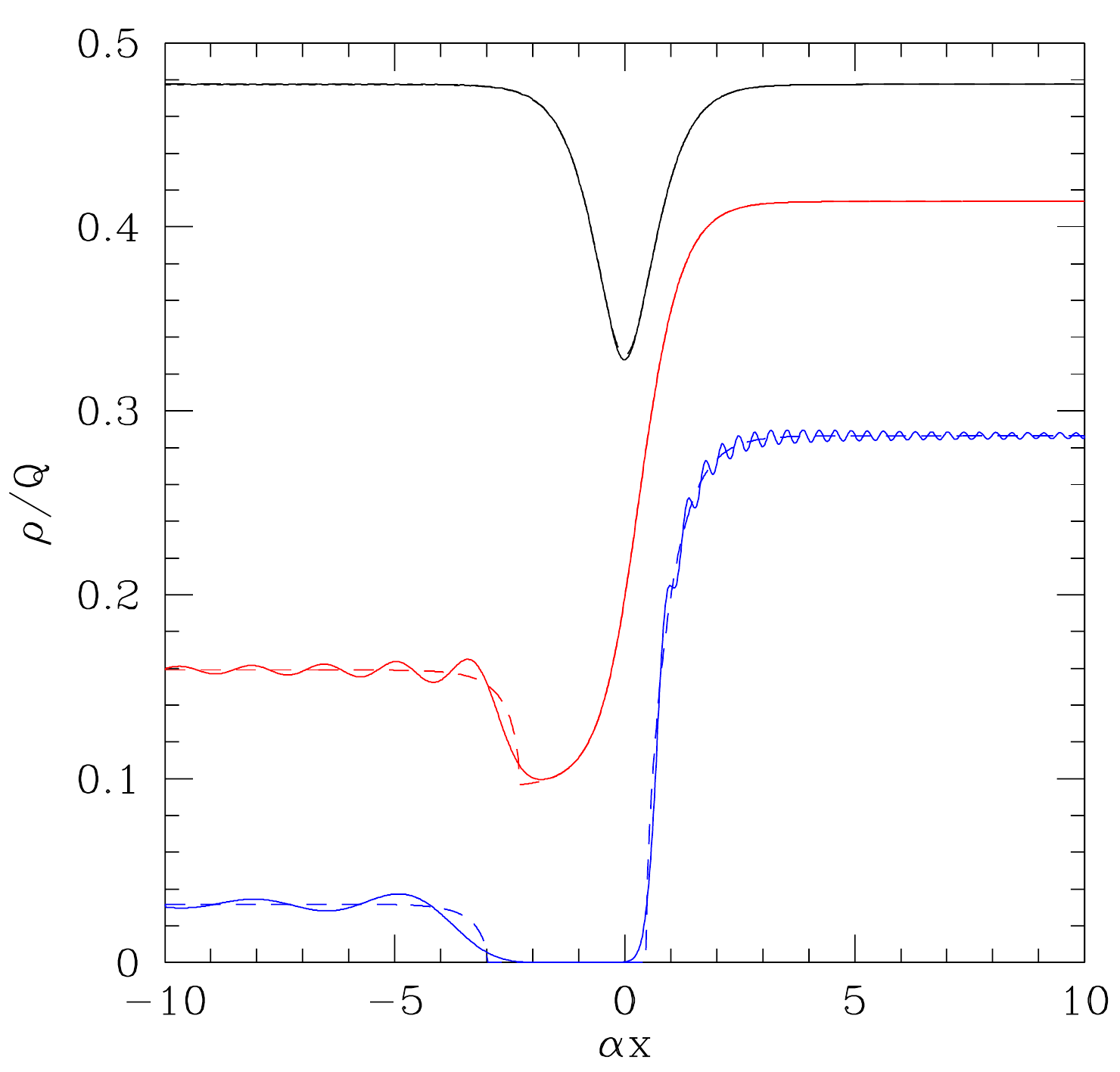}
\end{center}
\caption{Density profiles for three representative cases. The solid lines indicate the exact results for $\frac{\alpha}{Q} = 0.1$ while the dotted curves represent the asymptotic profiles. For the blue lines $k_F= 0.5\, Q$ and $k_0 = 0.4\, Q$. For the red lines $k_F= 0.9\, Q$ and $k_0 = 0.7\, Q$. For the black lines $k_F= 1.5\, Q$ and $k_0 = 0.4 \, Q$. In this last case, the asymptotic result is indistinguishable from the exact one.
}

\label{fig-dens}
\end{figure}

It is possible to obtain analytic expressions for the density profile and the velocity field in the $\frac{\alpha}{Q}\ll 1$ limit~\cite{cpt}, showing
the emergence of three physically different regimes according to the height of the barrier: $Q < k_-$, $k_-< Q < k_+$, $Q > k_+$.
A comparison between the exact density profiles, evaluated at $\frac{\alpha}{Q} = 0.1$, and the results in the limit $\frac{\alpha}{Q} \to 0$ is shown in Fig.~\ref{fig-dens}, for three representative values of $\frac{Q}{k_{F}}$.
\begin{figure} [h!]
\begin{center}
\includegraphics[scale=0.5]{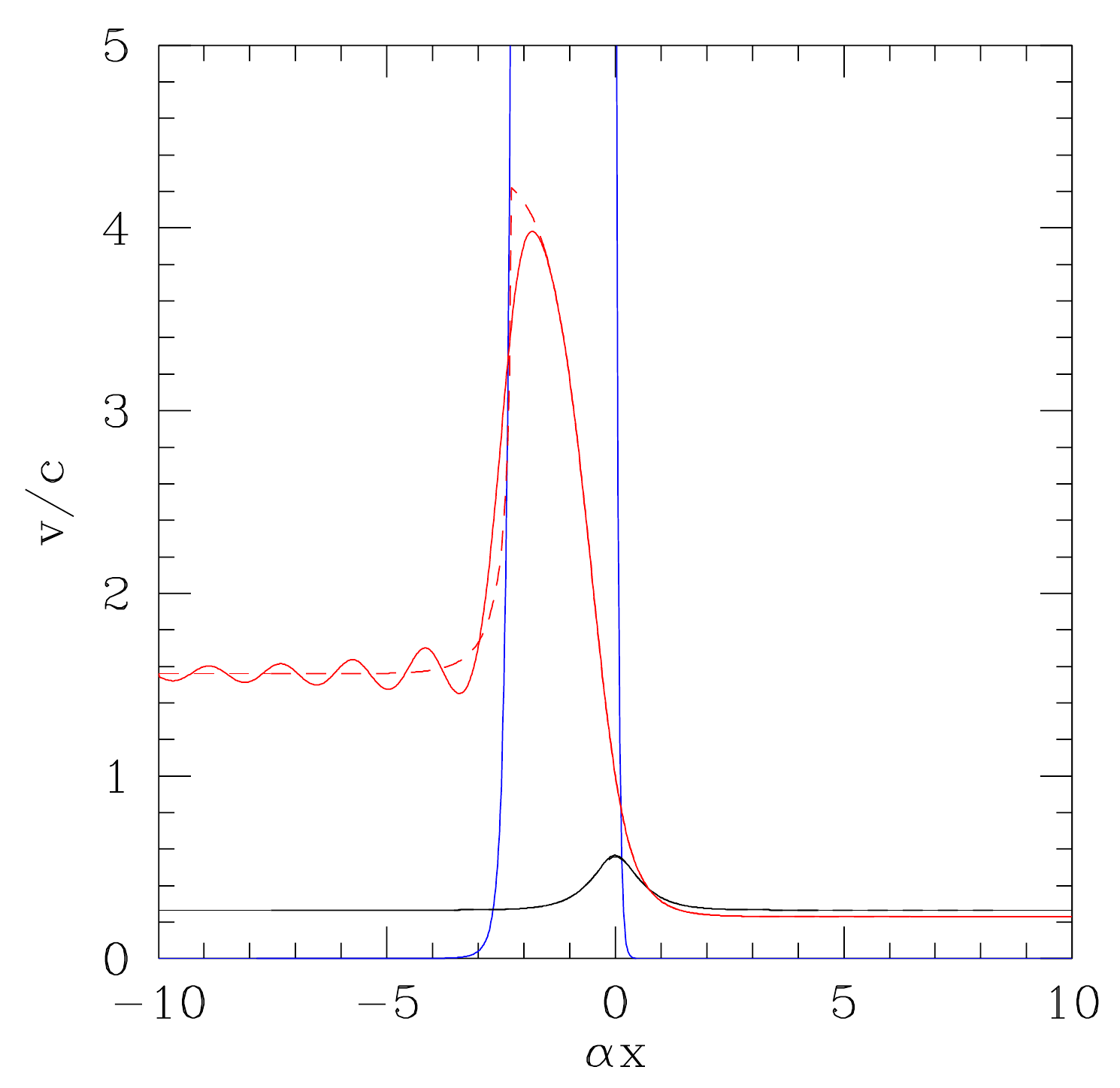}
\end{center}
\caption{Mach number for the same cases shown in Fig~\ref{fig-dens}. Note that in the case $Q> k_+$ the density drops abruptly in a large portion of the system, leading to very high $v/c$ ratios. 
}
\label{fig-mach}
\end{figure}
The Mach number:
\begin{equation}
\frac{v(x)}{c(x)} = \frac{|j|}{\pi\hbar\rho(x)^2}
\end{equation}
is shown in Fig.~\ref{fig-mach} for the same set of parameters. A supersonic region is present only if $Q >  k_-$, but for $Q>k_+$ the current $j$ vanishes for $\frac{\alpha}{Q} \to 0$. Note that in the semiclassical approximation  a sonic transition can be present only when the Mach number is monotonically increasing along the flow, while the exact solution allows for local maxima in $v/c$ as shown in Fig.~\ref{fig-mach}.

\begin{figure} [h!]
\begin{center}
\includegraphics[scale=0.5]{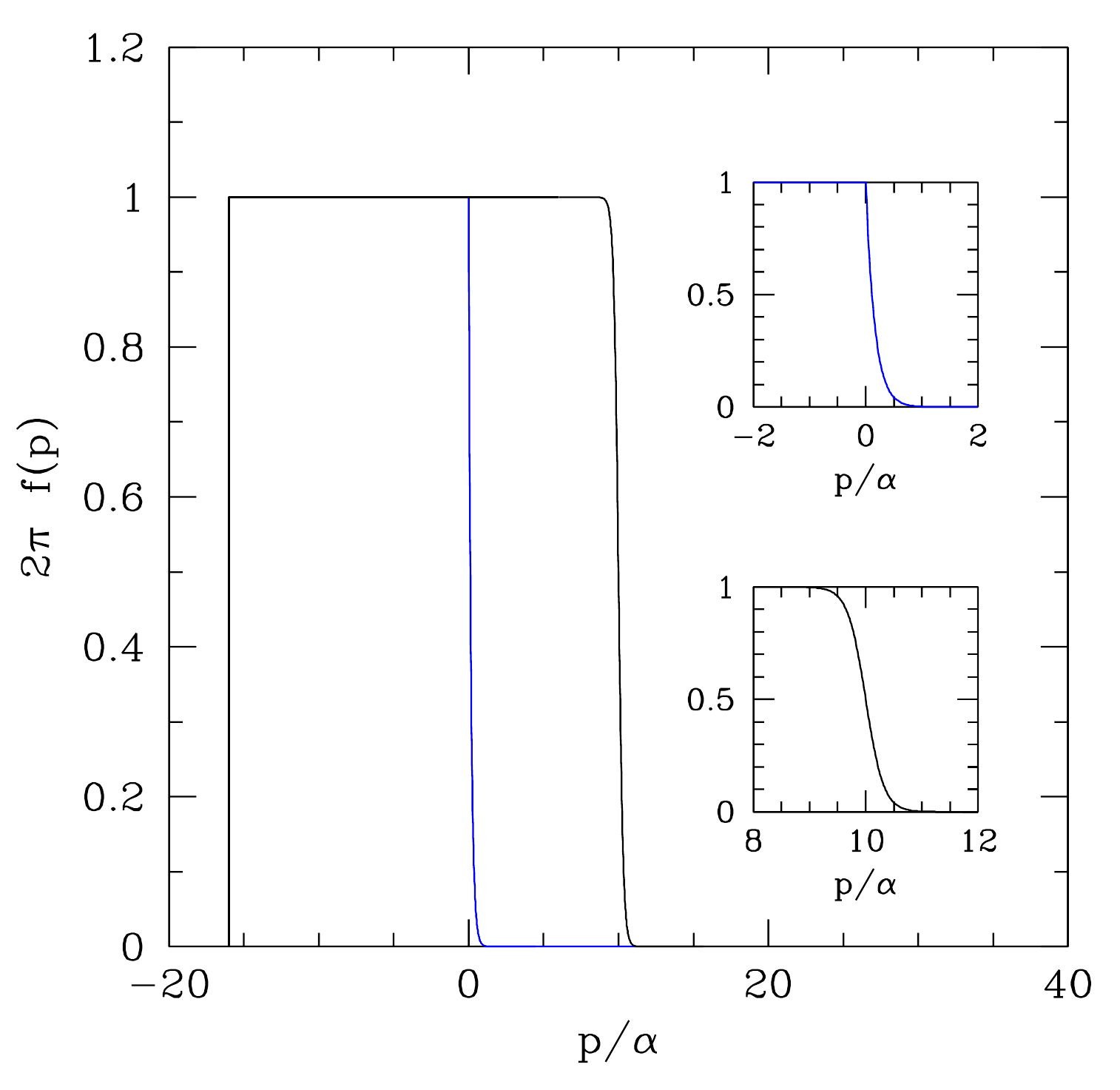}
\end{center}
\caption{Black line: effective Fermi distribution for a barrier defined by $\frac{\alpha}{Q}=0.1$ with 
parameters allowing for the presence of a sonic horizon ($k_F=0.9 Q$ and $k_0=0.7 Q$). For comparison,
the effective Fermi distribution for the case of a waterfall potential with the same parameters is also shown (blue line). 
In the insets a zoom of the two curves in the relevant ranges is reported. 
}
\label{fig-fp}
\end{figure}

The effective momentum distribution (\ref{effective}) can be also explicitly evaluated from the exact solution for any value of $\frac{\alpha}{Q}$: $f(p)$ is always affected by the presence of the barrier whenever the reflection coefficient is different from the classically allowed values $0$ or $1$ and sound waves propagate upstream even in the absence of a sonic transition but their amplitude is exponentially small in the parameter $\frac{Q}{\alpha}$ if the condition $k_- < Q < k_+$ is violated. Conversely, in this range, the ``phonon tail" of the momentum distribution has precisely the expected thermal character, as can be appreciated from Fig. \ref{fig-fp}. 
%and we can assess the validity of the relation~(\ref{unruh}).
In order to prove this statement, we focus on the form of the reflection coefficient, defining the effective momentum distribution~(\ref{effective}). In the $\frac{\alpha}{Q}\to 0$ limit, the exact expression for the transmission coefficient (\ref{tras}) 
simplifies, leading to: 
\begin{equation}
|R_p|^2 = 1-|T_p|^2 = \frac{1}{1+e^{\frac{2\pi}{\alpha}(p-Q)}} \, ,
\label{thermal}
\end{equation}
with corrections exponentially small in $\frac{Q}{\alpha}$ for $p\sim Q$.
Note that the inflection point of $|R_p|^2$ is located at $p=Q$, which belongs to the interval where the effective momentum distribution coincides with the reflection coefficient only if the condition $k_-< Q < k_+$ is satisfied. But the low ``temperature" behaviour of a Fermi gas is dominated precisely by the range of momenta close to the inflection point of $f(p)$ \cite{huang}: as a consequence, the effective distribution acquires a thermal character exclusively in this range of parameters. Then, by comparing Eq.~(\ref{thermal}) with the Fermi-Dirac distribution of a uniform gas, we can easily obtain the effective temperature; at low energy, in fact, the spectrum can be linearised close to $p\sim Q$: $\epsilon_p -\mu = \hbar v_F (p-Q)$ with $v_F=\frac{\hbar Q}{m}$ and $\mu=V_0$, leading to
\begin{equation}
k_BT_H = \alpha\,\frac{\hbar^2 Q}{2\pi m} = \frac{\alpha}{\pi Q}\,V_0 \, .
\label{hawking}
\end{equation}
It is clear from Eq.~(\ref{effective}) that no thermalization occurs near the left Fermi point $-k_+$, showing that only phonons flowing upstream (i.e., with positive momentum) are present at $x\to+\infty$. Thus, in the case of a barrier, if a sonic horizon is present and if the height of the barrier is smaller than the largest kinetic energy of the fluid particles, we detect a thermal flux of phonons flowing upstream in the $x \to +\infty$ region. This matches Hawking prediction~\cite{hawking2}. Furthermore, since the gravitational analogy subsists, we can test the agreement with Hawking result by comparing the temperature found with the one which would be obtained by means of Eq.~(\ref{unruh}). By use of the asymptotic expressions for $\frac{\alpha}{Q} \to 0$, it is easy to check that the sonic horizon is indeed present and is located exactly at the top of the barrier ($x=0$), where the analogue surface gravity $\kappa
=$ $\frac{d (c-v)}{d x}$ can be evaluated as $\alpha \frac{\hbar Q}{m}$, leading, via Eq.~(\ref{unruh}), to the expression~(\ref{hawking}), in full agreement with the analogue gravity picture.
This also shows that $\kappa \to 0$ in the asymptotic limit, implying that the relevant parameter $\kappa/\omega_{\text{max}}$, defying the domain of validity of the gravitational analogy, vanishes as $\frac{\alpha}{Q} \to 0$ satisfying the criterion established in Ref.~\cite{parentani1}. In fact, it can be shown that the full excitation spectrum, including the characteristic frequency $\omega_{\text{max}}$, remains finite in this limit~\cite{cpt}.

\section{The correlation function}

\begin{figure*} [ht!]
\subfloat[]{\includegraphics[height=5.2cm]{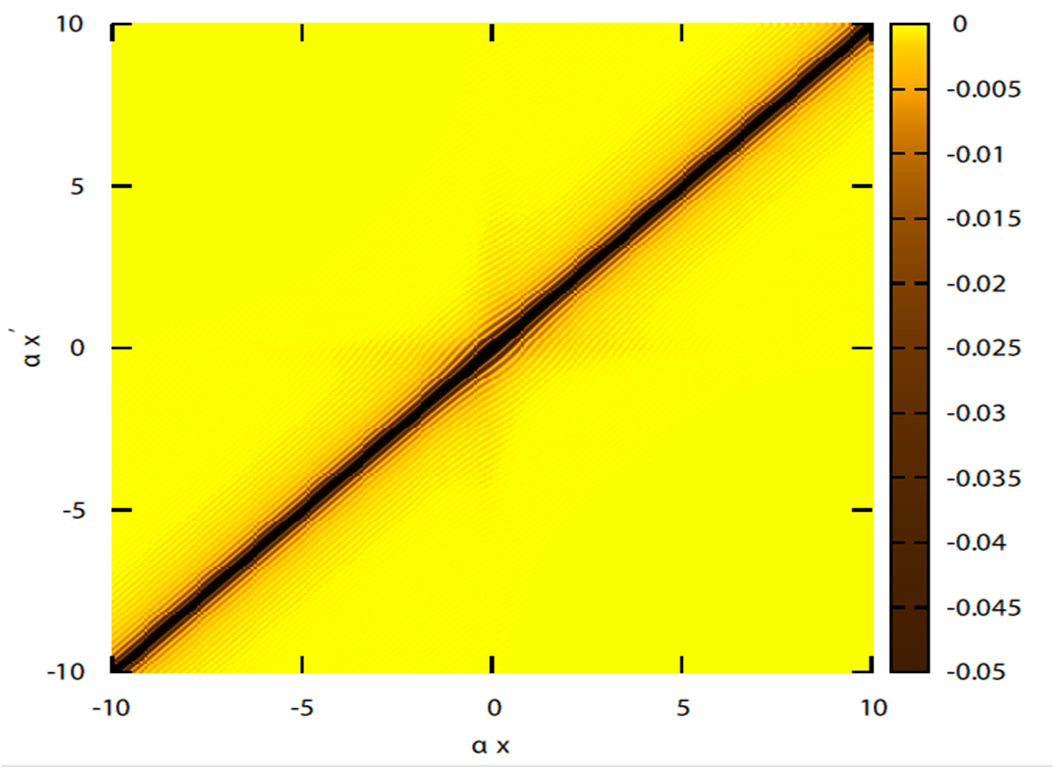}}
\subfloat[]{\includegraphics[height=5.2cm]{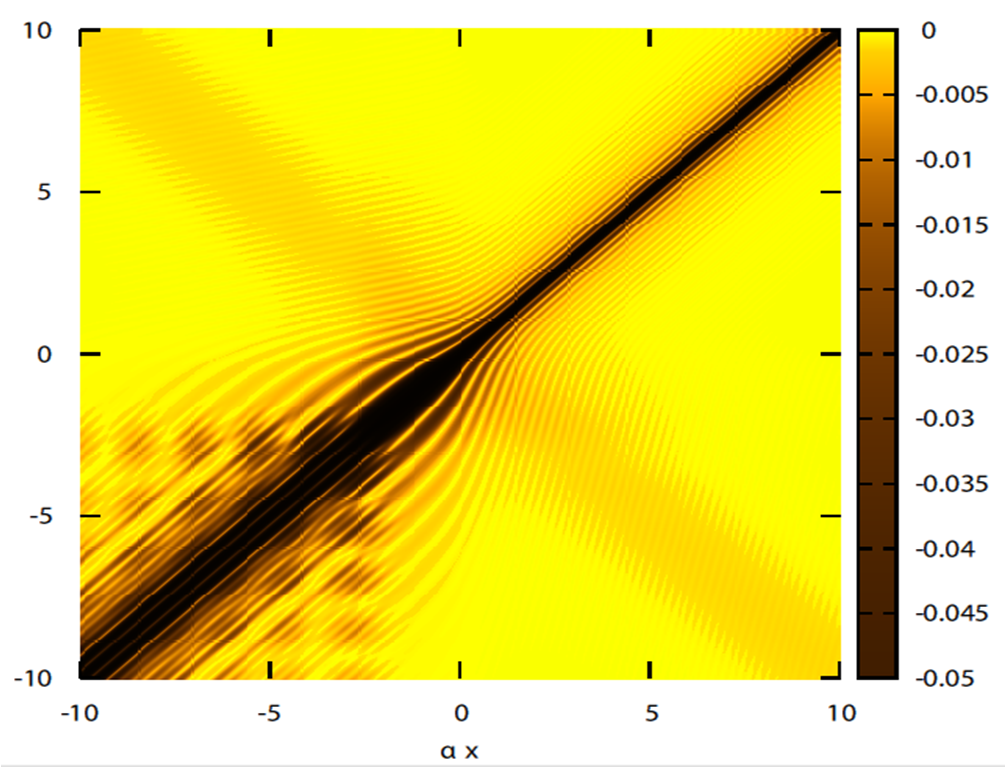}}
\subfloat[]{\includegraphics[height=5.2cm]{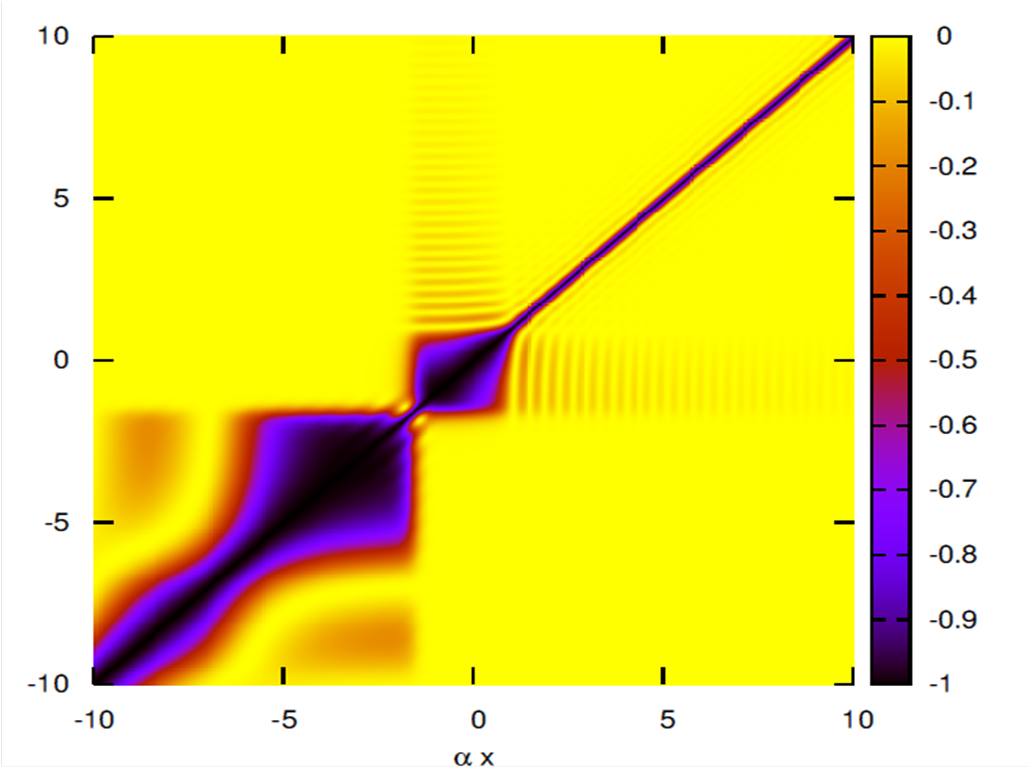}}
\caption{Colour plot of the correlation function~(\ref{dimcor}) in the three cases shown in Fig.~\ref{fig-dens}. From left to right: $k_F= 1.5\, Q$, $k_F= 0.9\, Q$, and $k_F= 0.5\, Q$. Note that the colour scale has been greatly enlarged in panel (a) and (b) in order to show the weak Hawking radiation signature, as mentioned in the text.}
\label{fig-corr}
\end{figure*}

Finally, let us evaluate the density correlation function, often referred to as the privileged marker of the presence of
Hawking radiation~\cite{iacopo,correlations}. The numerical evaluation of the connected two-point correlation function 
\begin{equation}
1+h(x,x^\prime)= \frac{\langle \sum_{i\ne j} \delta(x-\hat x_i)\,\delta(x^\prime-\hat x_j)\rangle}{\rho(x)\rho(x^\prime)}
\label{dimcor}
\end{equation}
is shown in Figs.~\ref{fig-corr}(a-c) in the three regimes of small, intermediate and large $\frac{Q}{k_{F}}$. 
Note that $h(x,x^\prime)$ takes values in the interval $[-1,0]$ but in panels (a) and (b) the scale of the pictures has been considerably enlarged to appreciate the small features present at $x+x^\prime\sim 0$. Besides strong correlations between nearby points $(x\sim x^\prime)$, a complex structure is present both in the upstream and downstream regions whenever a sonic transition takes place. However, the most interesting features are the weak correlations present only in the case $k_-< Q < k_+$ for $x+x^\prime \sim 0$, in agreement with the predictions of Ref.~\cite{iacopo,correlations}. These numerical results are confirmed by the analysis of the asymptotic behaviour of the solution for large $|x|$, $|x^\prime|$ and small $\frac{\alpha}{Q}$, where analytic expressions are available~\cite{cpt}. To lowest order in the small parameters, the connected correlation $h(x,x^\prime)$ can be written as the sum of a translationally invariant part $h_\pm(x-x^\prime)$ for $x\sim x^\prime \to \pm \infty$
and an off-diagonal term $h_{off}(x+x^\prime)$, given by:
\begin{equation}
%h_{off}(s) = -\frac{\alpha^2}{4}\,\frac{\cosh^{-2} \frac{\alpha s}{2}}{(3k_F-k_0-Q)(k_F+k_0+Q)}  
h_{off}(s) = -\frac{\alpha^2}{4}\,\frac{\cosh^{-2} \frac{\alpha s}{2}}{(k_+ +2k_--Q)(k_+ +Q)} \, .
\end{equation}
The off-diagonal correlations tend to a finite limit far from the potential only when the sonic horizon is present, but they are always very weak, being of order $\alpha^2\sim T_H^2$, while the diagonal part is of $O(1)$ in the $\frac{\alpha}{Q} \to 0$ limit. Note that off-diagonal correlations are present because both the reflection and the transmission coefficients acquire a non-vanishing value: a peculiar feature of quantum behaviour, absent in any classical system.

\section{Waterfall potential}

The results presented in this Letter refer to the special case of an external barrier of the form~(\ref{cosh}), where analytic expressions are available, but the same analysis can be performed for other potentials, corresponding to different experimental set-ups. Remarkably, in the case of a step (or ``waterfall") potential $V(x)=\frac{V_0}{2} [1+\tanh\alpha x]$, we find that the sonic horizon is pushed to $x_h\to+\infty$ as $\frac{\alpha}{Q} \to 0$ and, although sound waves are generated, the phonon tail of the effective momentum distribution is not thermal for any choice of the potential parameters~\cite{cpt} as exemplified in Fig. \ref{fig-fp}. Nevertheless, the two-point correlation function shows some off-diagonal features also in this case, in agreement with previous suggestions~\cite{parentani1}. 

\section{Conclusions}

In summary, the flow of a fluid past an obstacle is always accompanied by the emission of sound waves in the upstream direction, generally unrelated to the Hawking process. The exact solution of a microscopic model of a flowing Bose fluid has shown that the analogue Hawking effect requires additional conditions besides the occurrence of a supersonic transition in the flow: the obstacle must be a smooth repulsive barrier and the largest kinetic energy of the fluid particles must be bigger than the height of the potential. These conditions rule out other popular choices for the external potential, like the celebrated ``waterfall"~\cite{stein14,stein16}, and emerge as new requirements to be met in order to detect the analogue Hawking radiation in a quantum fluid. 
Indeed, it was already pointed out~\cite{liberati} that, if the smoothness condition is not met, finite energy phonons are excited, exposing corrections to the linear dispersion relation and introducing a finite lifetime for the elementary excitations. Now we have shown that, even in this case, an observer located far upstream would again detect a phonon flux. However, its energy spectrum would deviate from the predicted thermal distribution which characterizes Hawking radiation. This result has important consequences in the theoretical interpretation of experiments on flowing condensates.
When the previously summarized necessary conditions are satisfied, the thermal character of the analogue Hawking radiation follows directly from the exact solution and reflects in all (static and dynamic) density correlation functions evaluated in the far upstream region. For the parameter choice corresponding to the already-achieved experimental realization of the Tonks-Girardeau gas \cite{kinoshita1}, we can estimate a Hawking temperature of the order of several nano-Kelvin, a value which can be increased by tuning the scattering length or the radial trap frequency. 
Last, weak off-diagonal correlations in the two point function are present even in the limit of smooth potential. While this behaviour is expected in the case of a barrier - where the analogue Hawking radiation appears - it is also found for the case of the step potential, casting doubts on the effectiveness of this probe for the experimental demonstration of the occurrence of the analogue Hawking mechanism.
%In conclusion, the mere detection of sound waves emerging from a sonic horizon is not a sufficient condition to probe the occurrence of the Hawking mechanism in an analogue gravity experiment. 

\bibliographystyle{ieeetr}
\bibliography{Bibliography}

%\bibitem{b.a}
%  \Name{Author F., Author S. \and Author T.}
%  \REVIEW{Some Rev. A}{69}{1969}{9691}.
%
%\bibitem{b.b}
%  \Name{Author F. \and Author S.}
%  \Book{Some Book of Interest}
%  \Editor{A. Editor}
%  \Vol{9}
%  \Publ{Publishing house, City}
%  \Year{1939}
%  \Page{666}.
%
%\bibitem{b.c}
%  \Editor{Editor A.}
%  \Book{Some Book of Interest}
%  \Vol{9}
%  \Publ{Publishing house, City}
%  \Year{1939}
%  \Section{A}.

\end{document}